# HIGH TRANSPORT CRITICAL CURRENTS IN DENSE, METAL-CLAD SUPERCONDUCTOR WIRES OF $MgB_2$

S. Jin, H. Mavoori and R. B. van Dover

Bell Laboratories, Agere Systems/Lucent Technologies, Murray Hill, NJ 07974

Correspondence and requests for materials should be addressed to S.J. (e-mail: jin@lucent.com).

---------------------------------------------------------------------------

Technically useful bulk superconductors must have high transport critical current densities ($J_c$) at operating temperatures. They also require stabilization using a normal metal cladding to provide parallel electrical conduction, thermal stabilization, mechanical protection of generally brittle superconductor cores, as well as environmental protection especially for those containing reactive elements.

The recent discovery of the 39K superconductivity in the ceramic compound magnesium diboride ($MgB_2$)[1-5] not only re-ignited the enthusiasm for higher $T_c$ superconductivity in new classes of materials but also presented a new possibility for significant bulk applications, especially in view of the weak-link-free characteristics of the grain boundaries and the high magnetization critical currents observed. However, two important milestones, a demonstration of such a high critical current capability by actual transport measurements, and a feasibility demonstration of the required fabrication of metal-clad $MgB_2$ wires with high $J_c$, have not been achieved yet. Even the synthesis of moderately dense, bulk $MgB_2$ material itself with 39K superconductivity is a challenge[5] for many scientists and eventual industrial production because of the difficulties associated with the volatility and reactivity of magnesium, which often forces the use of rather inconvenient sintering heat treatment involving thousands of atmosphere pressure or magnesium vapor environment during heat treatment to get dense materials with acceptable stoichiometry.[1-4]

In this letter, we report the first successful fabrication of dense, metal-clad $MgB_2$ superconductor wires as well as the first demonstration of high transport $J_c$ of greater than 36,000 A/cm$^2$ in $MgB_2$. The fabrication technique that we devised uses only ambient pressure, yet produces very dense $MgB_2$ with little loss of stoichiometry. Such a technique can be useful for rapid and reliable compound syntheses and characterizations for future discovery of new or improved superconductors with higher $T_c$, higher Hc$_2$ or improved flux pinning characteristics. We also show that while inherently weak-link-free, the nature of the material is such that the grain boundaries of the polycrystalline $MgB_2$ can easily be altered to exhibit *induced weak-link behavior* with accompanying loss in critical current density by orders of magnitude.

$MgB_2$ is mechanically hard and brittle, and thus is not amenable to drawing into the desired fine wire geometry. We have made metal-clad wires by using the powder-in-tube technique similarly as in the case of the Y-Ba-Cu-O oxide superconductor.[6] The use of proper metal cladding has been found to be critical because of the strong chemical reactivity of $MgB_2$. Magnesium in $MgB_2$ tends to react and combine with many metals such as Cu or Ag to form solid solutions or intermetallics with lowered melting points[7] (as low as ~480$^o$C m.p. in the case of eutectic formation), which renders the metal cladding useless during sintering of $MgB_2$ at around 900 – 1000 $^o$C. Excluding elements with very low melting points such as Na or K, there are only a small number of essentially inert metals which exhibit no or little mutual solubility with Mg and do not form intermetallic compounds with Mg. These include Fe, Mo, Nb, V, Ta, Hf, W, and perhaps Ti (limited solubility for Mg). Of these, the refractory metals (Mo, Nb, V, Ta, Hf, W) are not ductile enough to endure substantial plastic deformation involved in fine wire or ribbon fabrication and tend to crack. Thus Fe and Ti appear to be the best candidate material as a practical clad metal or diffusion barrier for $MgB_2$ wire fabrication.

For $MgB_2$ wire/ribbon fabrication, we used an Fe tube (or a Cu tube lined with either Fe inner tube or Fe foil as a diffusion barrier). The Cu tube has an outside diameter (OD) of 6.35 mm. The inner Fe tube has an OD of 5 mm, a wall thickness of 0.5 mm, and was 10 cm long. One end of the tube was sealed by crimping and the tube was then filled, in argon atmosphere, with commercially available $MgB_2$ powder (98% purity, -325 mesh, procured from Alfa AESAR). The remaining end of the tube was also crimped by hand and the composite structure was then swaged (and wire drawn in some cases) to ~2-3 mm diameter rod (Fig. 1(a)) followed by cold rolling to a ribbon geometry with 0.25 – 0.5 mm thickness, 3 – 5 mm width, and ~60 cm length. The deformation involved in the present wire/ribbon fabrication process has an additional benefit of refining the $MgB_2$ grain size through pulverization of the powder inside the deforming metal clad and that of densely compacting the powder as well. For ease of handling , the ribbon was cut to ~5 – 10 cm long pieces and were given a sintering treatment in a laboratory furnace at 900$^o$C/30 minutes or 1000$^o$C/30 minutes in an argon atmosphere. A slow heating to the sintering temperature, lasting ~3 hours, was employed. For direct and reliable measurements of $T_c$ and $J_c$ from the superconductor, the metal cladding was mechanically removed. The bare $MgB_2$ ribbon so obtained was very dense giving an audible ping when dropped onto a hard surface or cut to smaller lengths using a hand-held wire cutter.

It is reassuring to find out that the bulk fabrication process we devised essentially maintains the intended stoichiometry of $MgB_2$. For example, the weight loss during sintering of the ribbons at 900$^o$C/30 min. in argon atmosphere, was measured to be only ~0.8% in terms of net $MgB_2$ weight change. By contrast, bare sintered pellets (prepared using a laboratory press at 3000 bar compression with the $MgB_2$ powder under argon atmosphere before and during the pressing), after the same 900$^o$C sintering, lost ~31% of weight from $MgB_2$ (equivalent to ~60% loss of magnesium!). The sintered pellet was porous, mechanically weak and very crumbly, thus not even allowing the electrical measurements, in agreement with Slusky et al's observation.[5]

Shown in Fig. 1(a)-(e) are the scanning electron microscopy (SEM) photomicrographs illustrating the structure of the heat treated (sintered) composite, metal-clad wire or ribbon of $MgB_2$. Figure 1(a) represents the sectional micrograph of the round preform composite, ~2 mm in dia., prior to cold rolling (sintered here for the purpose of metallography). Figure 1(b) shows the composite Cu/Fe/ $MgB_2$ ribbon wound into a solenoid configuration (~6 cm dia.) prior to sintering heat treatment, and Fig. 1(c) is the longitudinal cross-sectional micrograph from the final ribbon. It is seen that the superconductor core deforms continuously in conformation with the composite wire geometry during the swaging/wire drawing/rolling processes, presumably by particle-particle sliding. The Cu and Fe clad metal structure is well defined and distinguishable from the $MgB_2$ core which is ~35 μm thick. Figure 1(d) and Fig. 1(e) represent the high magnification microstructure of the 900°C and 1000°C sintered ribbons, respectively. A dense structure with an ultrafine grain size of ~1200 Å in average diameter is observed for the 900°C sample. This is much finer than the size of the starting $MgB_2$ powder material used (our SEM analysis gives an average of ~3 μm), indicating the occurrence of substantial grain refinement by the wire fabrication process. Such a grain refinement can allow the needed consolidation sintering at lower temperature thus reducing the extent of undesirable metallurgical reactions, such as the contamination of $MgB_2$ grain boundaries. A finer grain size in a weak-link-free superconductor could also be useful for flux pinning enhancement. The 1000°C sample, Fig. 1(e), exhibits an even denser microstructure with an appearance of some melting and crystallite formation. The average grain size of this sample is estimated to be ~2.5 times larger than that for the 900°C sample. Unlike the Y-Ba-Cu-O type superconductors[8,9], the larger grain size in $MgB_2$ does not lead to a significant increase in critical currents. We find that the magnetization $J_c$ and transport $J_c$ of the 1000°C sample are comparable to those for the 900°C sample.

The resistivity vs temperature curves for the bare $MgB_2$ samples stripped from the metal-clad ribbons were obtained by four point measurement using a 10 mA ac current. As shown in Fig. 2, a sharp superconducting transition occurred at $T_c$(onset) ~ 39.6K with the $T_c$ (mid point) being ~38.4K. The addition of a small amount (several percent) of excess Mg to the $MgB_2$ powder was found to slightly increase the $T_c$ as compared to the as-received $MgB_2$ alone which exhibited about 1- 2K lower $T_c$. The normal state resistivity $\rho(40K)$ was ~17 μΩ-cm with the $\rho(40K)/\rho(298K)$ ratio of ~1/2. These values are substantially different from those reported by Finnemore et al.[3] on high purity $MgB_2$ samples [$\rho(40K)$~1 μΩ -cm and $\rho(40K)/\rho(298K)$ ~20]. It is noted, however, that the presence of some impurities is not necessarily undesirable for superconductors and is often beneficial for obtaining higher $H_{c2}$ or enhanced $J_c$.

While high magnetization $J_c$'s on the order of $10^4$ ~ $10^5$ A/cm$^2$ have been reported for the bulk $MgB_2$ compound[2-4], such high critical current capability has not yet been confirmed by actual transport measurement. Larbalestier et al.'s samples[2] which gave an important clue on the strongly linked supercurrent current flow across high angle grain boundaries, yet yielded a transport $J_c$ of ~15 A/cm$^2$ due to the inhomogeneities in the sample. Canfield et al. reported transport $J_c$ of ~ 200 A/cm$^2$ in their MgB$_2$ samples made from boron whiskers.

We report here the first demonstration of high transport $J_c$ in the regime of $10^4$ ~ $10^5$ A/cm$^2$ on our MgB$_2$ ribbon samples. In order to avoid the complications with the presence of normal metal clad, the measurement was carried out with stripped (bare) MgB$_2$ ribbons with approximate dimensions of 0.1 – 0.3 mm thick x 1.2 mm wide x 13 mm long. Transport critical currents were obtained from V-I characteristic curves on passing a pulse current of 30 – 100 amperes from a capacitor bank[10] (~1 ms rise time and ~8 ms decay time) and monitoring with a transient digitizer. The current and voltage leads were attached to the sample using an In-10%Ag solder and ultrasonic soldering gun. The superconductor sample was either immersed in liquid He, or suspended above the liquid. A voltage criterion of 0.5 µV/mm was used to determine the critical current, primarily due to the limited voltage resolution in the digitizer. The samples typically experienced catastrophic failure on approaching or exceeding the critical currents due to the heating of contacts and lead wires, so only a lower limit to the critical current could be inferred. At 4.2 K, the measurement gave a lower limit $J_c$(transport) $> 3.6 \times 10^4$ A/cm$^2$ for the MgB$_2$ ribbons. At 20 K the lower limit of $J_c$(transport) was measured to be $2.3 \times 10^4$ A/cm$^2$. These values are consistent with the magnetization measurements, and represent the first explicit verification of the current-carrying capacity of MgB$_2$.

We have also measured the $J_c$(magnetization) for the similar bare ribbon samples and compared with the $J_c$(transport). The presence of circulating supercurrents in these samples was investigated by measuring the magnetization response of the samples using a vibrating sample magnetometer (VSM). Typical M-H curves for MgB$_2$ (900$^o$C samples) are shown in Fig. 3. The data measured at 4.2 K are particularly notable; the curve shows clear and sudden magnetization changes at fields below about µ$_0$H=0.5 T. Under these conditions, the sample is evidently subject to partial flux jumps[11] because of the large value of the critical current, a relatively large sample size (~2.5 x 2.5 x 0.3 mm), and small value of the heat capacity. This observation reemphasizes the need for superconductor stabilization using normal metal cladding. A crude interpolation of the data was employed to estimate $J_c$(magnetization) in the absence of the flux jumps (e.g., when they were suppressed by normal metal cladding) as indicated by the dashed line.

The M-H data can be interpreted using the Bean model[12] to yield the critical current as a function of field. For our square platelet samples, we use the formula $J_c=30\Delta M/W$, where $J_c$ is the critical current in A/cm$^2$, $\Delta M$ is the difference between the upper and lower branches of the M-H curve, in emu/cm$^3$, and W is the transverse width of the sample in cm. Figure 4 shows the critical current density [$J_c$(magnetization)] so inferred, as a function of magnetic field, for MgB$_2$ samples. The $J_c$(magnetization) values at 4.2K were ~3 x 10$^5$ A/cm$^2$ at H=0 and ~1 x 10$^5$ A/cm$^2$ at H=1T. At 20K, the values were ~1.2 x 10$^5$ A/cm$^2$ at H=0 and ~4 x 10$^4$ A/cm$^2$ at H=1T. The measured transport $J_c$ values (also plotted on Fig. 4) are comparable to the zero field $J_c$ (magnetization) values, and will most likely approach the $J_c$(magnetization) values upon future, better-controlled measurements with reduced contact resistance.

In view of the somewhat marginal upper critical field ($H_{c2}$) and flux pinning characteristics for applications of the $MgB_2$ superconductor at 20K or higher temperatures, it is important to improve such characteristics through structural or microstructural modifications. For example, chemical doping, introduction of precipitates or dispersoid particles, or atomic-scale control of defects such as vacancies, dislocations, grain boundaries, etc. may be attempted. In addition, possible effects of contamination of $MgB_2$ with other elements such as from the clad metal is of interest and needs to be understood.

As a part of such studies, we have evaluated the effect of several alloying metal elements on the critical current behavior of the $MgB_2$ material. 5 mole % each of fine metal particles of Fe, Mo, Cu, Ag, Ti (~1 – 10 μm average size), and Y (<~200 μm) were thoroughly mixed (using mortar and pestle) with the as received $MgB_2$ powder, and the metal-clad ribbons fabricated as described earlier. After sintering (900 ºC/30min.) and stripping off the clad metal, the $J_c$(magnetization) properties were evaluated as a function of field and temperature. It should be mentioned that during the ribbon fabrication process used here, the thickness of these metal particles also gets reduced by a factor of ~ 10.

Figure 5(a) shows the superconducting $T_c$ of $MgB_2$ samples with various metal powder additions measured as a.c. susceptibility vs temperature curves . Clearly the presence of these fine particles in $MgB_2$ during sintering at $900^oC$ have little effect on the $T_c$ of $MgB_2$ (except some broadening of the susceptibility transition in the case of Y which has a considerable mutual solubility with Mg, forms $566^oC$ eutectic, and reacts with B as well ). The same trend as in Fig. 5(a) was also confirmed with resistivity vs temperature measurements. This result suggests that the metal ions of Fe, Mo, Ag, Cu, Ti, and Y are not incorporated into the $MgB_2$ superconductor phase. This is not surprising in view of the observation[5] that many elements in the periodic table simply would not go into the small lattice space in $MgB_2$.

While the $T_c$ of $MgB_2$ remains unaffected by these elements, we notice a significant alteration of critical current behavior in these metal- containing $MgB_2$ samples as indicated in Fig. 5(b) ($J_c$ vs T curves) and in the lower part of Fig. 4 ($J_c$ vs H curves). The Fe addition appears to be least damaging while the Cu addition causes $J_c$ to be significantly reduced by 2-3 orders of magnitude with somewhat increased field dependence of $J_c$. Upon increasing the amount of added metal particles, e.g., to 20 mole %, even more severe degradation of $J_c$ properties is observed. This $J_c$ behavior resembles the well-known grain boundary weak link problem in the cuprate superconductors. We reconcile this observation by suggesting that the atoms of the added metal diffuse at the sintering temperature to the $MgB_2$ grain boundaries (or to the $MgB_2$ particle surface prior to the completion of the sintering reaction) and form a very thin reacted layer which is either nonsuperconducting or superconducting with reduced $J_c$, and which can be continuous or semi-continuous. The elements Cu, Ag, Y and Ti react with magnesium while Fe, Mo, Y as well as Ag, and Ti can react with boron to form intermetallics, presumably with different kinetics. We have not yet identified the nature or morphology of such layers in our preliminary SEM analysis. It is certainly possible that such a layer can be extremely thin, yet be very effective as grain boundary weak links since the superconducting coherence length in $MgB_2$ as well as the normal state coherence length in grain-boundary phases are on the order of 100 Ä or less. The presence of such grain boundary layers would strongly impede the flow of supercurrents from grain to grain. Thus, in view of the reactivity of Mg, any contamination of $MgB_2$ with foreign metal atoms needs to be carefully avoided or controlled in order to prevent such "*induced weak-link effect*". From these data, iron appears to be a particularly suitable material as one of the least weak-link-inducing clad metals or diffusion barrier metals for $MgB_2$ wire fabrication.

In summary, we report the first successful fabrication of dense, metal-clad $MgB_2$ superconductor wires as well as the first explicit verification of the transport current-carrying capacity of $MgB_2$ in the $J_c$ regime of $10^4 \sim 10^5$ A/cm$^2$. The ambient-pressure fabrication technique that we devised for synthesis of very dense $MgB_2$ with little loss of stoichiometry can be useful for rapid and reliable exploration for new or improved superconductors. We also show that the inherently weak-link-free $MgB_2$ can easily be altered to exhibit *induced weak-link behavior* with significant loss in critical current density.

**Figure Captions**

Fig. 1.  Photomicrographs showing (a) cross-sectional micrograph of round, swaged perform composite prior to cold rolling, (b) coil-shape wound ribbon of Cu/Fe/MgB$_2$, (c) longitudinal cross-sectional micrograph of Cu/Fe/MgB$_2$ ribbon, (d) microstructure of the 900$^o$C sintered MgB$_2$, and (e) microstructure of the1000$^o$C sintered MgB$_2$.

Fig. 2.  Resistivity vs temperature curve for the MgB$_2$ in the metal-clad wire.

Fig. 3.  M vs H  loops for the MgB$_2$ sample sintered at 900$^o$C.  Data taken at 4.2K show the effect of flux jumps at H<0.5 T, which occur only under conditions of extremely high critical currents and very low heat capacity and cause local thermal runaway.  At 35K, the M-H loop is completely closed for H>0.6T, that is, for H greater than the irreversibility field at this temperature.

Fig. 4.  Jc vs H plots for various MgB$_2$ samples.  Solid symbols indicate data inferred from M-H loops, as in Fig. 3.  The open symbols represent lower limits for the transport J$_c$  at 4.2K (upper data point) and at 20K (lower data point).

Fig. 5.  (a) Magnetic susceptibility vs temperature curves for MgB$_2$ containing fine metal particles.  (b) J$_c$ (magnetization) vs temperature curves.  Note that the additions of these metals do not affect the T$_c$ of MgB$_2$ much, yet the J$_c$'s are strongly decreased (except the case of Fe), apparently because of the *induced weak links* in the vicinity of MgB$_2$ grain boundaries.

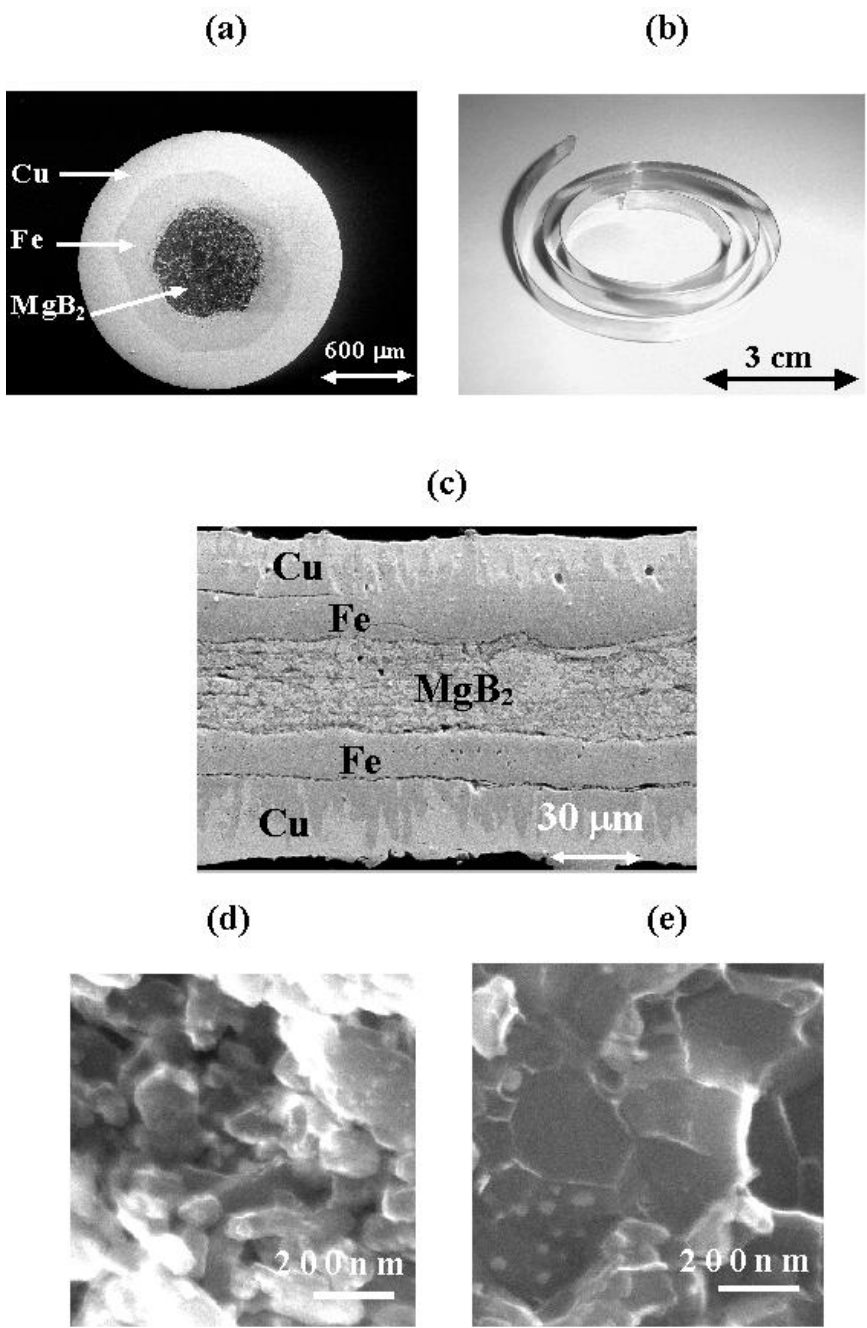

Fig.1
8

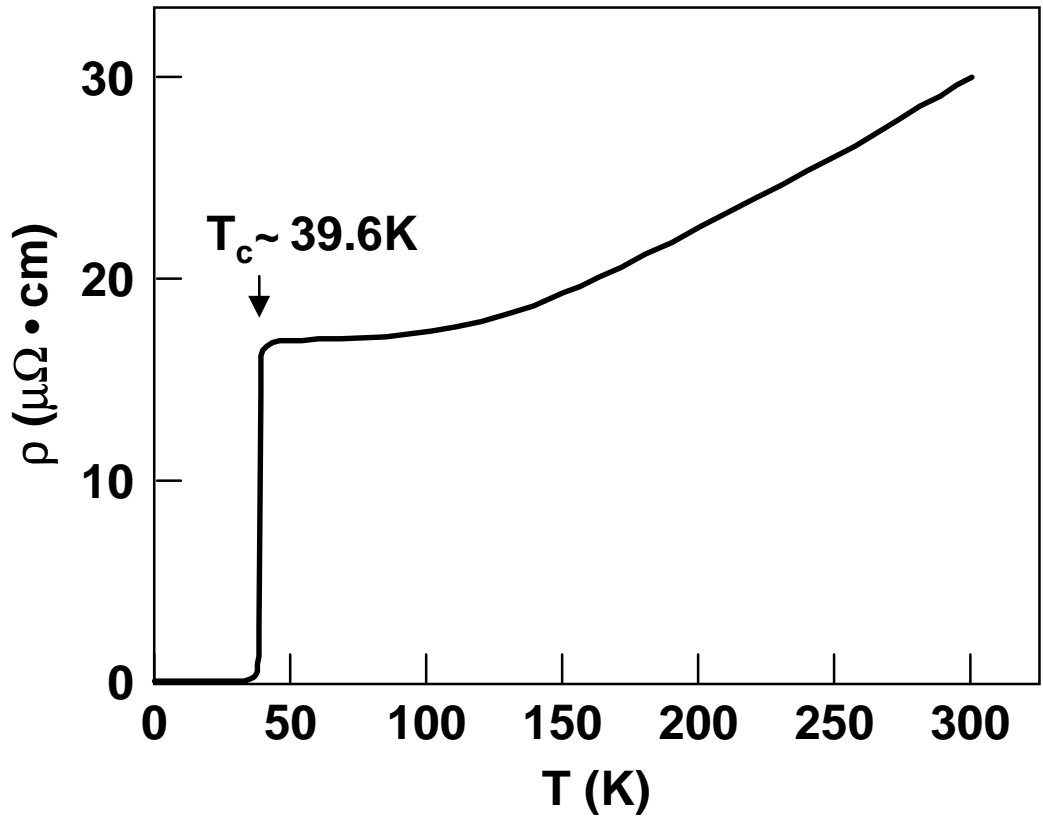

**Fig. 2**

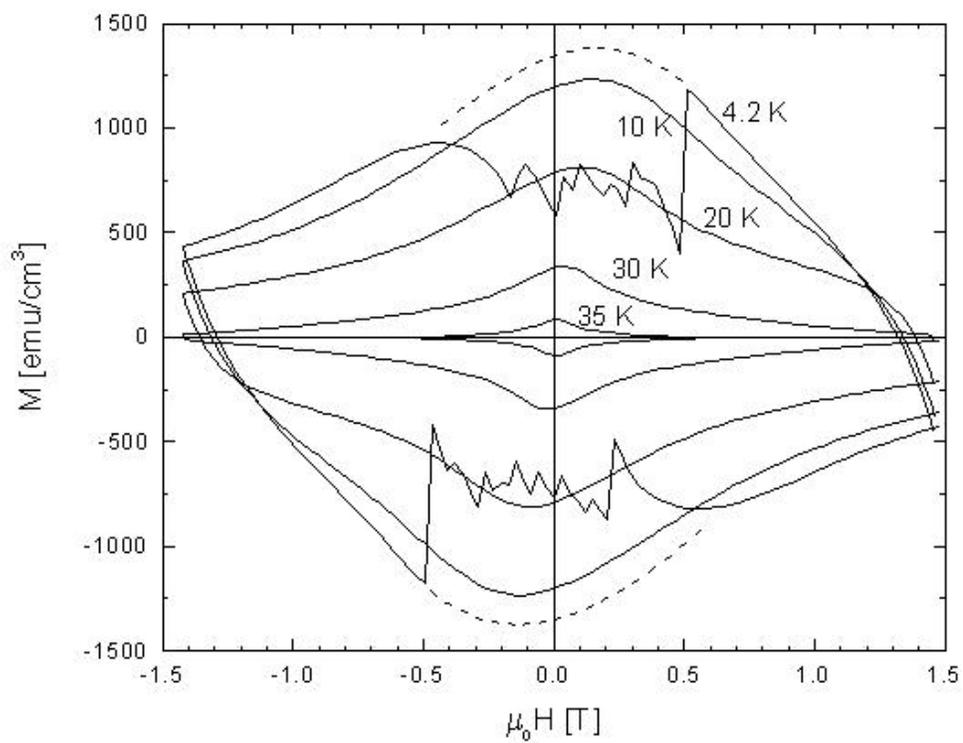

**Fig. 3**

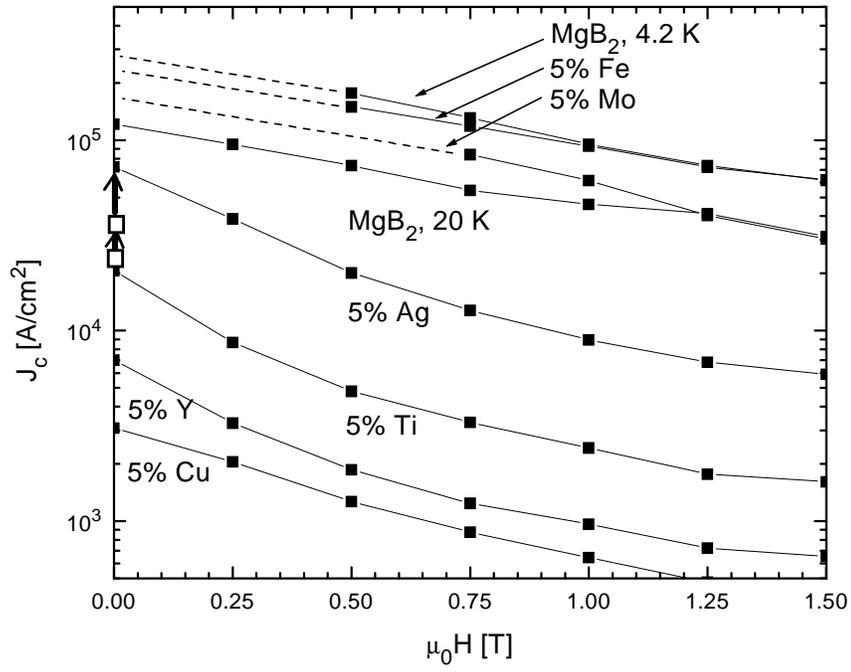

**Fig. 4**

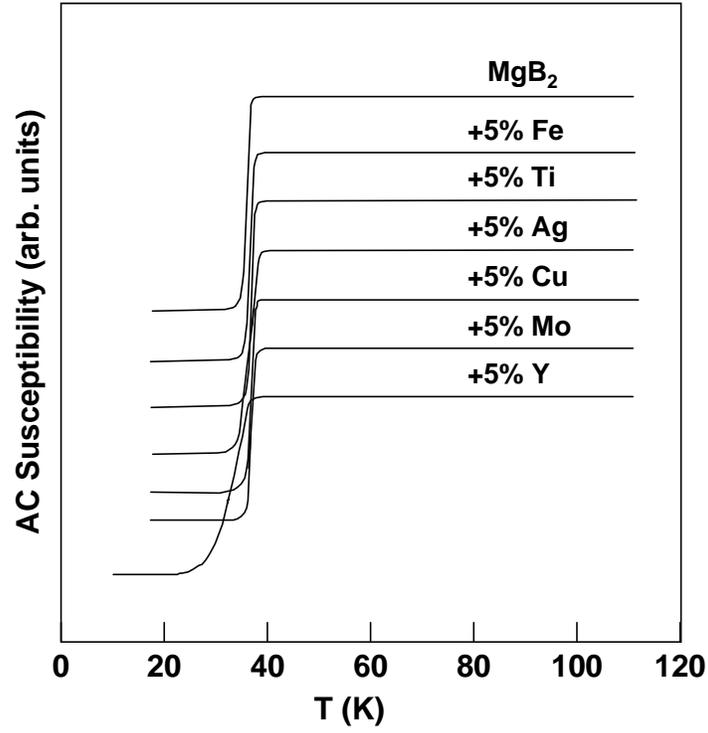

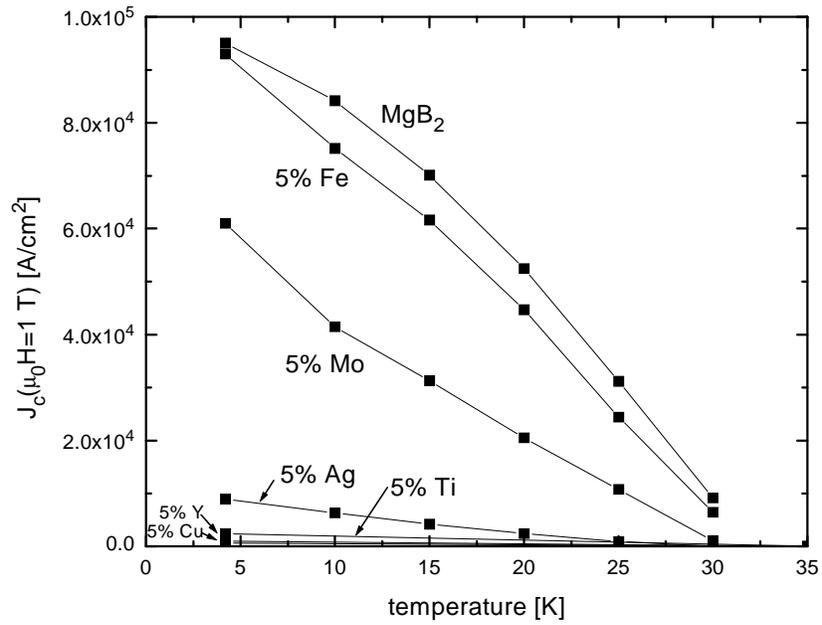

**Fig. 5**